# Topological polarization singularities induced by the non-Hermitian Dirac points

Jun Wang[1,†], Jie Liu[1,†], Peng Hu[1,2], Qiao Jiang[1], and Dezhuan Han[1,*]

[1]*College of Physics, Chongqing University, Chongqing 401331, China.*
[2]*College of Science, Chongqing University of Technology, Chongqing 400054, China.*
*†These authors contributed equally to this work.*
**e-mail*: dzhan@cqu.edu.cn

**Keywords**: non-Hermitian Dirac point, polarization singularities, exceptional points, circularly polarized states

**Abstract**

A Dirac point in the Hermitian photonic system will split into a pair of exceptional points (EPs) or even spawn a ring of EPs if non-Hermiticity is involved. Here, we present a new type of non-Hermitian Dirac point which is situated in the complex plane of eigenfrequency. When there is differential loss, the Dirac point exhibits a dual behavior: it not only splits into a pair of EPs with opposite chirality in the band structure but also induces a pair of circularly polarized states (C points) with opposite handedness in the far-field radiation. Furthermore, breaking the corresponding mirror symmetries enables independent control of these Dirac-point induced C points, facilitating the merging of two C points and generation of unidirectional guided resonances. Our results demonstrate an explicit relation between the band singularities and polarization singularities, and provide a new mechanism to generate unidirectional emission, which can be useful in the band engineering and polarization manipulation.

## 1. Introduction

The topological nature of optical field has been extensively investigated, unveiling a lot of intriguing physical phenomena, such as polarization singularities.[1-3] In real space, a conserved topological charge can be defined by considering the phase of a complex scalar field which is quadratic in the electric field. Examples of these singularities include circularly polarized states (C points) and vortex polarization singularities (V points). It has been demonstrated that the C points possess half-integer topological charges, whereas the V points carry integer topological charges. On the other hand, the polarization fields in momentum space have also been widely studied and polarization singularities are also revealed.[4-10] A typical example is the bound states in the continuum (BICs) in periodic photonic structures,[4-6] which can be interpreted as the vortex centers of polarization field and therefore the radiation to the far field is forbidden. Furthermore,

C points and their charge dynamics also exhibit many interesting properties in momentum space.[7-10] It is noted the topological charges in momentum space can be obtained by counting the winding times of polarization direction since the momentum space is normally two dimensional for photonic slabs.

This polarization field in momentum space originates from the non-Hermiticity of an open system. Distinct from the intrinsic loss and gain in PT-symmetrical systems,[11-15] it is radiation loss that plays the key role in this non-Hermiticity. Exceptional points (EPs), as topological band singularities, will appear if two or more eigenstates coalesce. It has been shown that a Dirac point in a closed system will split into a pair of EPs due to the radiation loss in the corresponding open system.[16, 17] The Fermi arc, connecting a pair of EPs, not only exhibits a nontrivial Berry phase along a closed loop encircling this arc, but also carries a half-integer topological charge in the far-field radiation.[10, 17] However, it is still a challenge to unravel the exact relationship between the band singularities and polarization singularities.

In this work, an explicit relation between the band singularities and polarization singularities is derived rigorously. A non-Hermitian Hamiltonian is adopted, in which a new type of Dirac point in the complex plane of eigenfrequency is formed by two orthogonal guided resonance (GR) modes with equal radiation loss. When the differential radiation loss of these two modes exists, this Dirac point will split into a pair of EPs with opposite chirality in the band structure, and meanwhile, induce a pair of C points with opposite handedness in the far-field radiation. The positions of the EP pair and C-point pair are obtained analytically. Intriguingly, the C-point pair induced by the Dirac point will move from one band to the other when the differential loss has a sign reversal. During this process, both the Berry phase of the wave function and the total topological charge of far-field radiation remain invariant. The C-point pair induced by the Dirac point can even be independently controlled by breaking the corresponding mirror symmetry and two C points induced by different Dirac points can merge again, leading to a new mechanism to generate unidirectional guided resonance (UGR) — the optical mode that radiates toward only one side of the structure.[8, 9, 18]

## 2. Theory based on a non-Hermitian Hamiltonian

Dirac points can exist in Hermitian photonic systems with some particular symmetries.[19-24] However, when a PhC slab with a finite thickness is considered, the radiation loss is inevitable and the system becomes non-Hermitian, therefore the Dirac point no longer exists and may split into an EP pair[16, 17] or an exceptional ring.[25, 26] Recently, the Dirac points in non-Hermitian systems have been demonstrated in a PT-symmetric case.[27, 28] To date, a non-Hermitian Dirac point in the complex plane of eigenfrequency is not reported for an open photonic system without intrinsic loss and gain. Here, we start by constructing such an accidental Dirac point with some Bloch wave vector **k** in a periodic system. The Hamiltonian of this non-Hermitian system can be written as:[29-31]

$$H = \Omega - i\Gamma, \tag{1a}$$

$$\Gamma = \frac{N}{2} D^\dagger D, \tag{1b}$$

where $\Omega$ and $\Gamma$ are the real and imaginary parts of Hamiltonian respectively, $D$ is the coupling matrix between the resonance modes and the radiation continuum with N identical open channels. If the system has an up-down symmetry, we have N = 2 and Eq. (1b) turns into $\Gamma = D^\dagger D$. In this Hamiltonian, we consider the case of two energy levels that correspond to two different resonance modes characterized by the resonance frequencies $\omega_{1,2}$ and radiation rates $\gamma_{1,2}$. The off-diagonal terms of $\Omega$ and $\Gamma$ correspond to the near-field and far-field coupling of the two modes respectively. Suppose the two resonances have a degenerate point at **k** = ($k_D$, 0) in the two-dimensional momentum space and the dispersion near this point is linear as $\omega_i = \omega_0 + c_i k_x$, where $k_x$ is the momentum displacement from $k_D$. Here, we assume that the two resonance modes are orthogonal along the $k_x$ axis, and their coupling strength is linearly proportional to $k_y$ when $k_y$ is small. The 2×2 coupling matrix $D$, which relates the two independent radiation channels with mutually orthogonal polarizations and the two resonances, can be written as:

$$D = i \begin{bmatrix} \sqrt{\gamma_1} & \delta_1 \\ \delta_2 & \sqrt{\gamma_2} \end{bmatrix}. \tag{2}$$

The off-diagonal term $\delta_i \propto k_y$ leads to the far-field coupling of the two modes. And for the near-field coupling, we also assume $\kappa$, the off-diagonal term of $\Omega$, is proportional to $k_y$. The Hamiltonian can be further expressed as $H = (\omega_0 - i\gamma_0)\mathbf{I} + \Delta H$, where $\omega_0 = (\omega_1 + \omega_2)/2$, $\gamma_0 = (\gamma_1 + \gamma_2)/2$,

$\Delta H = \Delta_0 \mathbf{I} + \Delta_i \sigma_i$, $\mathbf{I}$ denotes the identity matrix, $\sigma_i$ are Pauli matrices, $\Delta_i$ are the functions of $\Delta\gamma$ ($=\gamma_2-\gamma_1$) and $k_y$. More details are shown in Sec. 1 of Supporting Information.

When the differential loss $\Delta\gamma=0$ at $\mathbf{k} = (k_D, 0)$, $D$ is reduced to $i\sqrt{\gamma_0}\mathbf{I}$ and the Hamiltonian becomes $H_D = (\omega_0 - i\gamma_0)\mathbf{I}$, leading to a diabolic point at the complex frequency $\omega_0 - i\gamma_0$. The eigenfrequencies near this diabolic point are $\omega_\pm = \frac{1}{2}\left(H_{11} + H_{22} \pm \sqrt{f}\right)$, where $f = (H_{11} - H_{22})^2 + 4H_{12}^2$ is the discriminant of the characteristic polynomial, and $H_{ij}$ is the matrix element of $H$. As shown in Sec. 1 of Supporting Information, by expanding $f$ near the diabolic point, the real parts of eigenfrequencies are in fact linearly dependent on $\mathbf{k}$ as follows:

$$\omega_\pm \approx \frac{1}{2}\left((c_1 + c_2)k_x - 2i\gamma_0 \pm |c_1 - c_2|k\sqrt{1 + \tau\sin^2\theta}\right). \tag{3}$$

The eigenvectors near this diabolic point are $|\psi_\pm\rangle \propto [\psi_1, 1]^T$, where $\psi_1 = \left(H_{11} - H_{22} \pm \sqrt{f}\right)/2H_{12}$. The right and left eigenvectors of this non-Hermitian Hamiltonian satisfy the following equations: $H|\psi_\pm^R\rangle = \omega_\pm|\psi_\pm^R\rangle$, $H^\dagger|\psi_\pm^L\rangle = \omega_\pm^*|\psi_\pm^L\rangle$. The Berry phase[32, 33] around this diabolic point can be calculated as:

$$\gamma_B = i\oint_C \langle\psi_\pm^L|\nabla_\mathbf{k}|\psi_\pm^R\rangle \cdot d\mathbf{k} = \pi. \tag{4}$$

This non-trivial Berry phase gives rise to a Dirac cone. The Dirac cone can be classified into type-I and type-II according to the dispersion around the Dirac point.[34] When $c_1=-c_2$ and $\tau=0$, $\omega_\pm=\pm|c_1|k-i\gamma_0$ which corresponds to a conventional Dirac cone. When the signs of c1 and c2 are opposite, it will form a type-I Dirac cone. In contrast, when the signs of $c_1$ and $c_2$ are the same, the Dirac cone is of type-II.

This Dirac cone will be deformed when a mass term is introduced.[16, 25] If the introduced mass term is purely imaginary, the band crossing will be remained but the Dirac point transforms into the Fermi arc. The two eigenfrequencies, both their real and imaginary parts, coincide at the EPs at which the discriminant $f=0$. Compared to the Dirac point, the eigenvectors no longer form an orthogonal basis but become parallel at the EPs which can be denoted by $|\psi_{EP}\rangle = (\pm i, 1)/\sqrt{2}$, where $\pm$ represent the states with different chirality. The coordinates of EPs, $\mathbf{k}_{EP}$, can be solved from $f=0$ and expressed as a power series of $\Delta\gamma$. To the zero order of $\Delta\gamma$, $\mathbf{k}_{EP}^{(0)} = 0$ because the EP pair will merge into the Dirac point at $\Delta\gamma = 0$, and therefore $\mathbf{k}_{EP}$ can be expressed by $\mathbf{k}_{EP}^{(1)} + O(\Delta\gamma^2)$, where the first-order term is solved as follows (see details in Sec. 2 of Supporting Information):

$$k_{x,\,EP} \approx \sqrt{\gamma(d_1+d_2)\,\Delta\gamma/c_3(c_1-c_2)} \qquad (5)$$
$$k_{y,\,EP} \approx \pm\Delta\gamma/2c_3$$

The far-field radiation can be related to the eigenstates through the $D$ matrix as: $S^- = (E_x,\,E_y)^T = D|\psi_\pm\rangle$.[31] The topological charge of a polarization singularity is defined by $q = \frac{1}{2\pi}\oint_L d\mathbf{k}\cdot\nabla_{\mathbf{k}}\phi(\mathbf{k})$, where $L$ is a closed loop in momentum space, $\phi(\mathbf{k})$ is the orientation angle of polarization ellipses. The C points, as a special type of the radiation states with half-integer topological charges, can also be represented by $(\pm i,\,1)/\sqrt{2}$ [+/−: left-/right-handed circular polarization (LCP/RCP)]. These states have the same form as that of the EPs but completely different physical meaning, the entries of the former correspond to two orthogonal polarizations, while the latter are the components of wave function.

The correspondence between EPs and C points can be easily observed if the D matrix relating them is proportional to a unit matrix. In fact, the D matrix near the Dirac point can be expressed as $D = i(D_0 + \Delta D)$, where $D_0 = \sqrt{\gamma_0}\mathbf{I}$, $\Delta D = D_i\sigma_i$. To obtain the coordinate of C points, $\mathbf{k}_{CP}$, the eigenvectors near the EPs can be expanded as follows:

$$|\psi_\pm\rangle \approx \frac{1}{\sqrt{2}}\begin{bmatrix}\pm i \\ 1\end{bmatrix} + \frac{\delta}{\sqrt{2}}\begin{bmatrix}\mp i \\ 1\end{bmatrix}. \qquad (6)$$

From $S^- = D|\psi_\pm\rangle$, it is obvious that the zeroth order $D_0|\psi_{EP}\rangle$ gives rise to the radiation state of a C point, which indicates clearly that the lowest order of $\mathbf{k}_{CP}$ and $\mathbf{k}_{EP}$ are equal. The first-order correction gives that $\sqrt{\gamma_0}\delta \pm iD_1 - D_3 = 0$ and the next order of $\mathbf{k}_{CP}$ is found to be of $\Delta\gamma^2$, i.e.

$$\mathbf{k}_{CP} = \mathbf{k}_{EP}^{(1)} + O(\Delta\gamma^2). \qquad (7)$$

It can demonstrate that $\mathbf{k}_{CP}$ and $\mathbf{k}_{EP}$ have the same second order corrections, and their difference exists in the third-order terms. More details are shown in Sec. 2 of Supporting Information. We also note that the signs ± in Eq. (6), namely the different chirality of EPs, correspond to the far-field states of LCP and RCP exactly since the D matrix will not flip the signs ±. In this way, we prove that, with the introduction of $\Delta\gamma$, a Dirac point can not only split into a pair of EPs with opposite chirality in the band structure, but also induce a pair of C points with opposite handedness in the far-field radiation.

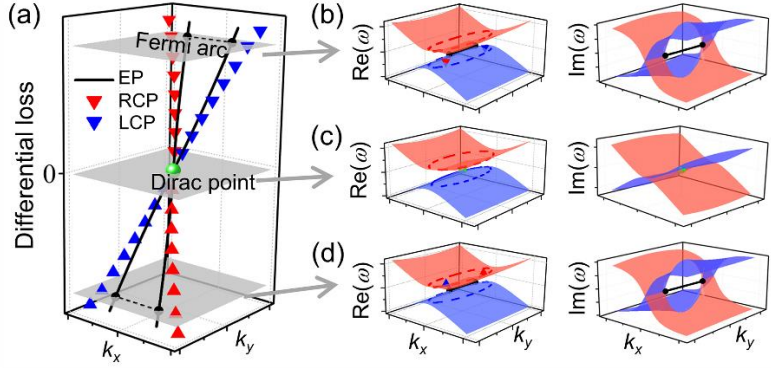

**Figure 1.** A Dirac point induce an EP pair in the band structure and a C-point pair in the far-field radiation. (a) Schematic of the loci of Dirac-point induced singularities in momentum space when the differential loss $\Delta\gamma$ varies. (b-d) Three examples of band structures for different $\Delta\gamma$, showing the merging of an EP pair and a C-point pair at the Dirac point. The red (blue) dashed lines represent the isofrequency contours on the upper (lower) bands.

In Fig. 1(a), the loci of singularities in momentum space are shown as functions of $\Delta\gamma$, and three examples of band structures for different $\Delta\gamma$ are shown in Figs. 1(b)-(d). The Dirac point exists when $\Delta\gamma=0$. When $\Delta\gamma$ is finite, an EP pair in the band structure as well as a C-point pair in the far-field radiation arise from this Dirac point. The pair of EPs (black dots) are connected by a Fermi arc (black lines) in the band structures, and the pair of C points with opposite handedness but the same topological charge emerge near the EPs, marked by the red (RCP) and blue (LCP) triangles. It should be emphasized that the coordinates of the EPs and the C points have the same linear behavior near the Dirac point since they have the same first-order term of $\Delta\gamma$. And the differences of their coordinates are manifested in the three- and higher-order terms.

The nontrivial Berry phases of the wave function on the upper and lower bands remain invariant even though the Dirac point deforms into the Fermi arc. It is significant that whether the total topological charges of far-field radiation are still conserved in the presence and absence of the C-point pair. When $\Delta\gamma=0$, any isofrequency contours on the lower and upper band encircling the Dirac point [Fig. 1(c)] carry the same half-integer topological charge. When $\Delta\gamma$ becomes finite, the local behavior near the Fermi arc is significantly altered, leading to the emergence of a C-point pair. However, the far-field polarization states along a large closed contour encircling the Fermi arc only undergo small and continuous variations, thus maintaining the conservation of the total topological charge (see details in Sec. 2 of Supporting Information). It is noted that both the Fermi arc and C points contribute to the total charge, indicating an interdependence between their individual charges. In Fig. 1(b), the isofrequency contour (blue dashed circle) on the lower band encircling both the Fermi arc and two C points carries a half-integer charge as the same as that on the upper band (red dashed circle). The conservation of topological charge results in the fact that the charge of the Fermi arc has to be opposite in sign for the two bands.[10] In addition, the C-point pair will cross the Dirac point and move from one band to the other when $\Delta\gamma$ has a sign reversal. In this process, the topological charges of the Fermi arc on the two bands also have a sign reversal, which is guaranteed by the conservation of total topological charge.

## 3. Examples in the system of photonic crystal slabs

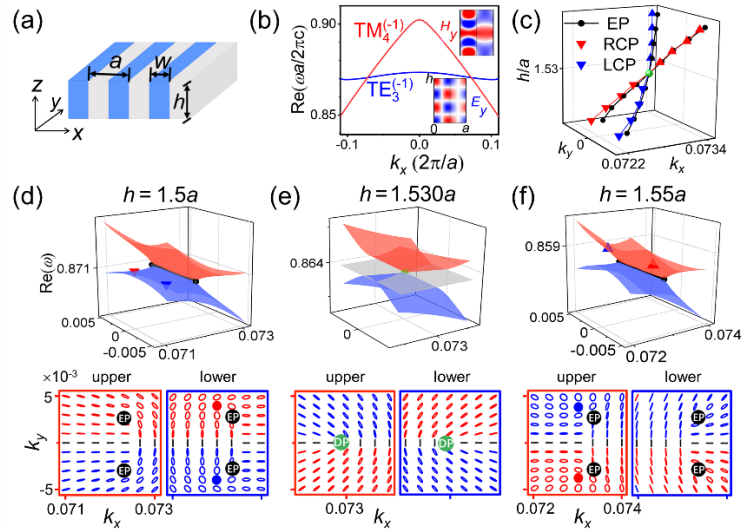

**Figure 2.** A type-II Dirac point splits into an EP pair and induces a C-point pair. (a) Schematic of a PhC slab. (b) Simulated dispersion of $TE_3^{(-1)}$ (blue line) and $TM_4^{(-1)}$ (red line) bands along the $k_x$ axis. The field distributions at the Γ point are shown in the insets. (c) The loci of Dirac-point induced singularities in momentum space when $h$ varies. (d) The band structure for $h = 1.5a$. A C-point pair emerge on the lower band. The polarization states are shown in the lower panel in which the ellipses indicated in red (blue) correspond to right-handed (left-handed) states, and the red (blue) dots represent the states with RCP (LCP). (e) A type-II Dirac point for $h = 1.530a$. The equal-energy surface at this Dirac point is shown in gray. (f) The band structure for $h = 1.55a$. The C-point pair moves to the upper band.

To verify the theory above, one-dimensional (1D) PhC slabs are taken as an example. Figure 2(a) shows the PhC slab which has a period $a = 400$ nm in the $x$ direction, uniform in the $y$ direction, and a finite thickness $h$ in the $z$ direction. The PhC slab consists of alternating dielectric ($\varepsilon=4$) and air ($\varepsilon = 1$) layers, each with a width of $w = 0.5a$. All the GR modes in this slab can be classified into the TE-like and TM-like ones and labeled by $GR_i^{(j)}$ which represents the $i$th guided mode with $j$ being the band index in the reduced-zone scheme. Numerical simulations are performed and Fig. 2(b) plots the band structures of $TE_3^{(-1)}$ (blue) and $TM_4^{(-1)}$ (red) along the $k_x$ axis. The Dirac point is found at $\mathbf{k}_D = (0.073, 0)$ $(2\pi/a)$ for $h = 1.530a$. The corresponding band structure forms a type-II Dirac cone in momentum space, a degenerate point with tilted conical dispersion [Fig. 2(e)]. The polarization states of the far-field radiation for the upper and lower bands are also shown below the band structures. It can be observed that the topological charge of polarizations around the Dirac point is 1/2 both on the upper and lower bands. The emergence of EPs and C points are dependent on the differential loss, which can be achieved by varying the thickness $h$ here. As $h$ varies away from $1.530a$, this type-II Dirac point splits into an EP pair in the band structure and induces a C-point pair with opposite handedness but equal topological charge $q = 1/2$ in the far-field radiation. The loci of both the band and polarization singularities are shown in Fig. 2(c), indicating that they are derived from the Dirac point clearly. When $h$ decreases from $1.530a$, the pair of C points emerge on the lower band [Fig. 2(d)]. However, when $h$ increases, this pair of C points move to the upper band [Fig. 2(f)]. During this process, the total topological charge $q = 1/2$ is conserved for both the two bands, and the topological charges of Fermi arcs for these two bands have a sign flip when the C-point pair moves from one band to the other. The simulated results can be well described by a non-Hermitian Hamiltonian and confirm our theory above (see Sec. 4 of Supporting Information).

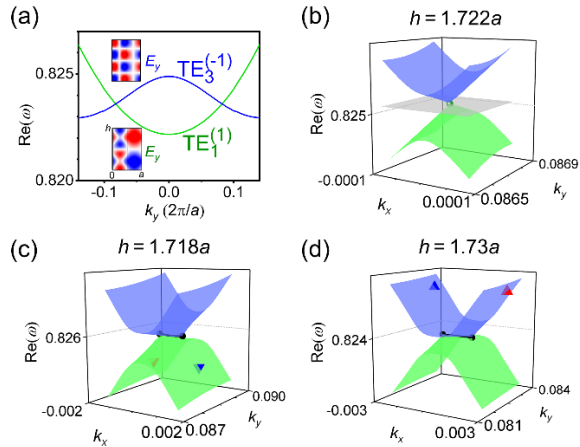

**Figure 3.** EP pair and C-point pair induced by a type-I Dirac point on the $k_y$ axis. (a) The band structures of

$TE_3^{(-1)}$ (blue line) and $TE_1^{(1)}$ (green line) along the $k_y$ axis. (b) A type-I Dirac point for $h=1.722a$. The equal-energy surface at the Dirac point is shown in gray. (c) Band structure for $h=1.718a$. A C-point pair exist on the lower band. (d) Band structure for $h=1.73a$. The C-point pair moves to the upper band.

The existence of Dirac point and its induced singularities are ubiquitous if the two-level Hamiltonian consists of two photonic modes with orthogonal radiation polarization. We further show a type-I Dirac point on the $k_y$ axis. Fig. 3(a) plots the band structures of $TE_3^{(-1)}$ (blue) and $TE_1^{(1)}$ (green) along the $k_y$ axis for $h=1.722a$. The far-field polarizations of these two bands are orthogonal on the $k_y$ axis although they are parallel on the $k_x$ axis. Their band structures form a type-I Dirac cone [Fig. 3(b)], where the isofrequency contour collapses into a single point at $\mathbf{k}_D = (0, 0.0867)$ $(2\pi/a)$. Similar to the results in Fig. 2, the EP pair in the band structure and C-point pair in the far-field radiation are induced by this Dirac point, and the conservation of total topological charge is preserved when $h$ varies. The polarizations of the far-field radiations are shown in Sec. 5 of Supporting Information.

## 4. A new mechanism to generate unidirectional guided resonances

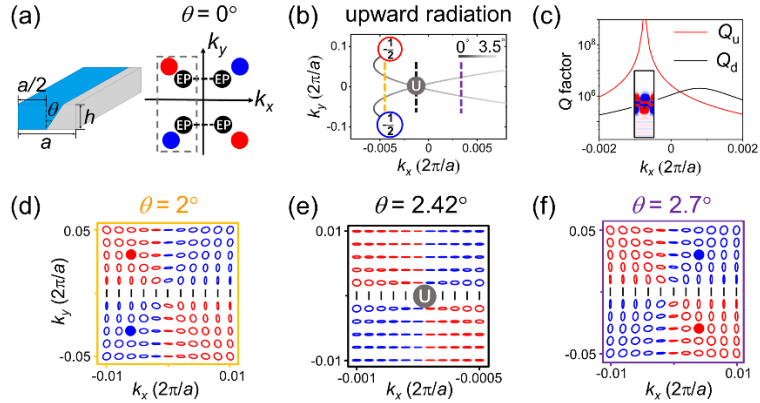

**Figure 4**. Unidirectional guided resonance (UGR) arising from the Dirac-point induced C points. (a) Schematic of the tilted sidewall of 1D PhC slab and the positions of singularities for $\theta=0°$. $\theta$ denotes the tilt angle. (b) The loci of two C points [red and blue dots in the dashed box of (a)] induced from different Dirac points as $\theta$ increases from $0°$ to $3.5°$. The UGR (denoted by "**U**") is generated at $\theta=2.42°$. (c) $Q$ factors for the upward ($Q_u$) and downward ($Q_d$) radiation. (d)-(f) The polarization states of upward far-field radiation at $\theta= 2°$, $2.42°$, and $2.7°$.

The EP pair and C-point pair will merge into the Dirac point in the up-down symmetric systems. However, the C-point pair can be even independently manipulated by breaking the corresponding mirror symmetry. Here, we further show that the two C points induced by different Dirac points can also merge together and generate a UGR when the $\sigma_x$ and $\sigma_z$ mirror symmetries are broken.

The symmetry breaking can be implemented by tilting the sidewall as shown in Fig. 4(a), in which $\theta$ represents the tilt angle. The right panel of Fig. 4(a) shows the positions of singularities when $h = 1.7a$ and $\theta = 0°$, where the two C points (red and blue dots) in the dashed box are associated with different Fermi arcs that cross the positive and negative $k_y$ axes. The loci of these two C points for the upward radiation are shown in Fig. 4(b) when $\theta$ varies. The far-field polarization at $\theta = 2°$, 2.42°, and 2.7° are showed in Figs. 4(d)-(f) respectively. It can be observed that the two C points move closer to each other when $\theta$ increases from 0° and they merge into a V point at $\theta = 2.42°$ where the upward radiation is forbidden and a UGR is formed. Because of the breaking of up-down symmetry, the $Q$ factor ($= \omega/2\gamma$) can be decomposed into $Q_u$ (upward) and $Q_d$ (downward) which are shown in Fig. 4(c). It can be seen the $Q_u$ (red) diverges near $k_x = -0.0007(2\pi/a)$ whereas $Q_d$ (black) remains finite. The electric field profile ($E_y$) of this UGR is shown in the inset of Fig. 4(c), indicating this GR mode has only the downward radiation.

## 5. Conclusion

In this work, we showed the existence of Dirac points with complex eigenfrequencies in non-Hermitian photonic systems, and proved the explicit connection between the band singularities and far-field radiation singularities, that is, a Dirac point can not only split into an EP pair in the band structure but also induce a C-point pair in the far-field radiation. The relation between the Dirac point, EPs, and C points are derived in a mathematically rigorous way. Therefore, the Dirac points, both type-I and type-II, and their induced C points are universal in the photonic systems. The C-point pair can be manipulated and moved from one band to another. Furthermore, when the up-down symmetry of the structure is broken, the C points associated with the different Fermi arcs can merge together and generate a V point in momentum space, providing a new mechanism to generate UGRs. Our results may be useful for band engineering and polarization manipulations in optical systems.

## 6. Methods

On the non-Hermitian Dirac Hamiltonian, behavior of EPs and C points near the Dirac point, and the related coupled mode theory, the derivations and discussions are given in the Supporting Information in detail. The eigenmodes of photonic crystal slabs are calculated by using a finite-element method. Periodic boundary conditions are applied in the *x* and *y* directions, while the scattering boundary condition is applied in the *z* direction. The polarization states of the far-field radiation are obtained from a Fourier transform of the electromagnetic field in the far field.

## Supporting Information

Supporting Information is available from the Wiley Online Library or from the authors.

## Acknowledgments

We would like to express our sincere gratitude to Professors Z. Q. Zhang and C. T. Chan for their insightful discussions throughout the course of this research. We also thank Professors L. P. Yang, L. Shi and J. Zi for helpful discussions. This work was supported by National Natural Science Foundation of China (12074049, 12147102).

## Conflict of interest

The authors declare no conflicts of interest.

## Data availability

The data that support the findings of this study are available from the corresponding author upon reasonable request.